\documentstyle[11pt]{article}
\let\tsection\section
\renewcommand{\section}{\setcounter{equation}{0}\tsection}

\begin{document}
\begin{center} 
CURRENT FLUCTUATIONS IN THE ONE DIMENSIONAL SYMMETRIC EXCLUSION PROCESS
WITH OPEN BOUNDARIES
\vskip10pt

B. Derrida\footnote{
 Newton Institute, 20 Clarkson Road, Cambridge, CB3 00EH, UK and
Laboratoire de Physique Statistique,
Ecole Normale Sup\'erieure, 24 rue Lhomond, 75231  Paris Cedex 05, France
(permanent address)},
B. Dou\c{c}ot\footnote{Laboratoire de Physique Th\'eorique et des Hautes
Energies,
Universit\'e Denis Diderot, 4 Place Jussieu, 75252 Paris Cedex 05,
France}
  and P.-E. Roche\footnote{Centre de Recherches sur les Tr\`es Basses
Temp\'eratures CNRS, 25 avenue des Martyrs, 38042 Grenoble Cedex 9,
France (permanent address) and Laboratoire de Physique 
 de la Mati\`ere Condens\'ee
de l'Ecole 
 Normale Sup\'erieure, 24 rue Lhomond, 75231  Paris Cedex 05, France}
\vskip10pt
submitted to Journal of Statistical Physics
\\
September 30, 2003
\end{center}
\vskip20pt
\noindent {\bf Abstract}
We calculate the first four cumulants of the integrated current of the
one dimensional symmetric simple exclusion process of $N$ sites with open boundary conditions.
For large system size $N$, the generating function of the integrated current  depends on the densities $\rho_a$ and $\rho_b$ of the two reservoirs and on the fugacity $z$, the parameter conjugated to the integrated current, through a single parameter.
Based on our expressions for these first four cumulants, we make a
conjecture  which leads to a  prediction for all the higher cumulants.
In the case $\rho_a=1$ and $\rho_b=0$, our conjecture gives    the same    universal distribution  
as the one obtained by Lee, Levitov and Yakovets
for one dimensional quantum conductors in the metallic regime. 

\newpage

\section{Introduction\label{introduction}}
The study of the current through a system in contact with two reservoirs
at unequal chemical potentials or at unequal temperatures is one of the
most studied aspects of the theory of non-equilibrium systems \cite{LLP,BLR}.

For the last decade, there has been an increasing interest in the study     
of the fluctuations of the current of quantum particles (fermions)  through a disordered wire.
It is now well established that the quantum statistics of the particles determines the distribution of the fluctuations
of the current, and that in the  metallic regime \cite{BeB,BlB}, this
distribution  \cite{LLY,Naz,GGM} is universal.  
More recent works have shown that the main property of the quantum nature of the particles 
which was responsible for these universal fluctuations is the Pauli exclusion
principle \cite{N0,DB,N,RD,GG}. 

Here we consider the symmetric simple exclusion process  SSEP
\cite{Ligg,KOV,HS,ZS,SS}
which  is a stochastic model of classical particles 
with hard core interactions 
(and without inertia) which diffuse
on a finite chain with open boundary conditions. The chain is  in contact
at its two ends with two reservoirs of particles at unequal densities
\cite{DEHP,DLS1,DLS2,BDGJL,BDGJL2}.
The combined effects of the stochastic injection  and removal of particles at the two boundaries
and of  the diffusive nature of the hard core particles produce a fluctuating current.
We calculate the first four cumulants of the integrated current.
Based on our results for these four cumulants, we  give a conjecture  
for all the higher cumulants and for the whole distribution of the current fluctuations.      

The  fluctuations of the current in exclusion processes  is also a subject with  a long history \cite{R,BKT,FF,DL,PS}.
Most of the known results obtained so far concern infinite geometries
\cite{R,BKT,PS} or systems with periodic boundary conditions
\cite{DL,LK,DA,PPH} (see \cite{DEMal} for the variance of the integrated current of the asymmetric simple exclusion process with open boundaries).

Our paper is organised as follows: in section 2 we define the model and we summarize our results.
In section 3, we show how the first two cumulants can be calculated from the steady state properties.
In section 4, we write a hierarchy (see also Appendix \ref{hierarchy}) for the  correlation functions 
on which our approach is based. In sections 5 and 6 we solve this hierarchy, in a low density expansion, where at each order the hierarchy can be truncated.
Appendix \ref{GCRa} gives a derivation 
 of the Gallavotti-Cohen relation {\cite{GC,LS} for
the SSEP with open boundaries. Appendix \ref{ruin} points out the analogy with
multi-particle ruin problems.
\section{Definition of the model  and main results}
\subsection{The symmetric exclusion process with open boundaries}
In the one dimensional symmetric  simple exclusion process, each site  $i$ (with $1 \leq i \leq N$) of a one dimensional lattice
of $N$ sites is either occupied by a single particle or empty.  A configuration ${\cal C}$  at time $t$ is therefore fully determined by $N$  binary variables  $\tau_i(t)$,  the occupation numbers of the $N$ sites ($\tau_i(t)=1$ if site $i$ is
occupied and $\tau_i(t)=0$ if site $i$ is empty).
In the bulk, each particle  independently  attempts to jump to its right neighboring site, and to its left neighboring site, in each case at rate $1$. It succeeds if the target site is empty; otherwise nothing happens
(this means that during time $t$ and time $t+dt$ with $0 < dt \ll 1$, a particle at site $i$ jumps to site $i-1$
with probability $(1 - \tau_{i-1})dt$,   to site $i+1$
with probability $(1 - \tau_{i+1})dt$  and does not move with probability
$ 1 - (2 - \tau_{i-1} - \tau_{i+1})dt$).
At the left boundary
 particles are injected at site $1$ at rate $\alpha$
and removed from site $1$ at rate $\gamma$. Similarly  at the right
boundary, particles are removed from  site $N$ at rate $\beta$ 
and injected at site $N$ at rate $\delta$.

For general values of $\alpha, \beta, \gamma,\delta$, a current of particles flows through the system and we want to study the
 fluctuations of this current. To do so, we denote by
 $Q(t)$ the number of particles which have moved from the left
reservoir into the system during time $t$ (so $Q(t)$ is the number of
particles which have jumped into the system  at site $1$ minus the number of particles
which have left the system from site $1$).
We want to calculate the distribution of the total charge $Q(t) $ during a long time $t$.

For finite $N$ the system has $2^N$ internal configurations $ {\cal C}$ (each site can be either occupied  by a particle or empty).
Let $p_t({\cal C})$ be the probability of finding the system in
configuration ${\cal C}$ at time $t$.
As the dynamics is a Markov process, the evolution of the probability
$p_t({\cal C})$ of finding the system in configuration ${\cal C}$ at time
$t$ can be written as
\begin{equation}
{d p_t({\cal C}) \over dt} = \sum_{\cal C'} [W_1({\cal C},{\cal C'}) +   W_0({\cal C},{\cal C'})+
 W_{-1}({\cal C},{\cal C'})] p_t({\cal C'}) 
\label{evo1}
\end{equation}
where we have decomposed the Markov matrix into three parts, depending on whether when  the system jumps from  configuration ${\cal C'}$ to  configuration ${\cal C}$,
$Q(t)$ increases by $1,0$ or $-1$.
(the matrix $W_0$ contains all the diagonal terms which are all negative as well as all the non-diagonal elements corresponding to moves
which do not take place at the left boundary, i.e. do not change $Q(t)$).
One 
way to determine the distribution of $Q(t)$ is to calculate its generating function
$   \langle z^{ Q(t)} \rangle $.

If we define $P_t({\cal C},Q)$ the probability that the system is in configuration ${\cal C}$
at time $t$ and that $Q(t)=Q$, one has
\begin{eqnarray}
{d P_t({\cal C},Q) \over dt} =  \sum_{\cal C'} W_1({\cal C},{\cal C'}) P_t({\cal C'},Q-1) +   W_0({\cal C},{\cal C'}) P_t({\cal C'},Q)
&& \nonumber  \\
+ W_{-1}({\cal C},{\cal C'})
  P_t({\cal C'},Q+1) &&
\label{evo2}
\end{eqnarray}
Then the generating functions ${\cal P}_t({\cal C}, z)$ defined  by
\begin{equation}
{ \cal P}_t({\cal C}, z) = \sum_{Q=-\infty}^\infty  P_t({\cal C},Q)   \ z^Q
\label{PCzdef}
\end{equation}
 satisfy
\begin{equation}
{d {\cal P}_t({\cal C},z) \over dt} =  \sum_{\cal C'}  \left[ z  \  W_1({\cal C},{\cal C'})  +   W_0({\cal C},{\cal C'}) + {1 \over z} W_{-1}({\cal C},{\cal C'}) \right]
  {\cal P}_t({\cal C'},z)
\label{PCz}
\end{equation}
If we introduce the matrix $M_z$ defined by
\begin{equation}
M_z({\cal C},{\cal C'}) = z  \  W_1({\cal C},{\cal C'})  +   W_0({\cal C},{\cal C'}) + {1 \over z} W_{-1}({\cal C},{\cal C'}) 
\label{Mz}
\end{equation}
it is clear from (\ref{PCz}) that
in the long time limit
\begin{equation}
   \langle z^{ Q(t)} \rangle = \sum_{\cal C}{ \cal P}_t({\cal C}, z)  \sim e^{\mu \  t } 
\label{mudef}
\end{equation}
where $\mu$ is the largest eigenvalue of the matrix $M_z$.
So this largest eigenvalue $\mu$ fully determines the distribution of
$Q(t)$ in the long time limit \cite{DL}.
\subsection{Symmetries of  $\mu$}
In principle, $\mu$ depends on six parameters: the input rates $\alpha, \beta, \gamma,\delta$ at the two boundaries, the fugacity $z$
and the number of sites $N$.
There are three symmetries in the system that leave $\mu$ unchanged:
\begin{enumerate}
 \item {\it The left-right symmetry}: if we exchange the roles of  $\alpha,\gamma$ and  $\delta, \beta$, this has the effect of 
exchanging the roles of the left and of the right boundaries, and so the statistical properties of $Q(t)$ are replaced by those of
$-Q(t)$.
Therefore $\mu$ should satisfy
\begin{equation}
\mu( \alpha,\gamma, \delta,\beta,z,N) =
\mu( \delta,\beta, \alpha,\gamma,{1 \over z},N) 
\ .
\label{left-right} 
\end{equation}

\item {\it The particle-hole symmetry}: instead of counting the number of
particles $Q(t)$ entering at the left boundary, one can as well count the
number $-Q(t)$ of holes entering at the left boundary. Now the holes are
injected  at rate $\gamma$ and removed at rate
$\alpha$ at the left boundary and they are injected at rate $\beta$ and
removed at rate $\delta$ at the right boundary. They also jump with the
same exclusion rules as the particles in the bulk. Therefore, this symmetry implies
that 
\begin{equation}
\mu( \alpha,\gamma, \delta,\beta,z,N) =
\mu( \gamma,\alpha, \beta,\delta,{1 \over z},N)  \ .
\label{particle-hole}
\end{equation}
\item {\it The Gallavotti-Cohen symmetry}: the rates $ \alpha,\beta,\gamma, \delta$ 
represent the transfer of particles between the system and reservoirs 
at densities
$\rho_a= {\alpha \over \alpha + \gamma} $ and $\rho_b = {\delta
\over \beta + \delta} $
at the two boundaries  (sites $1$ and $N$).
When $\rho_a=\rho_b$, the system is in equilibrium and the dynamics satisfy detailed balance with respect
to a Bernoulli measure \cite{DLS2}  at density $\rho=\rho_a= \rho_b $.
One can always think that the case  $\rho_a \neq \rho_b $ represents the effect of an external field which
enhances the flux of particles from one reservoir into the system, a
situation for which (as explained in the Appendix \ref{GCRa})  the Gallavotti-Cohen relation holds \cite{GC,LS}.
This implies that
\begin{equation}
\mu( \alpha,\gamma, \delta,\beta,z,N) =
\mu( \alpha,\gamma, \delta, \beta,{\gamma\delta \over \alpha \beta z},N)
\ .
\label{Gallavotti-Cohen}
\end{equation}
  
\end{enumerate}
\subsection{Main results }
When $N$ is large, one  finds, at least pertubatively
in powers of $\alpha$ and $\gamma$ (see sections 5 and 6) that 
$\mu$ depends only on the densities $\rho_a$ and $\rho_b$ of the left and
right reservoirs  
\begin{equation}
\rho_a = {\alpha \over \alpha + \gamma} \ \ \ ; \ \ \ \rho_b = {\delta
\over \beta + \delta} 
\label{rhoa-rhob-def}
\end{equation}
instead of the four parameters $\alpha,\beta,\gamma$ and $\delta$.

The three symmetries (\ref{left-right})-(\ref{Gallavotti-Cohen})
 then become
\begin{equation}
\mu( \rho_a,\rho_b,z,N) =
\mu( \rho_b,\rho_a,{1 \over z},N) 
\label{left-right-bis}
\end{equation}

\begin{equation}
\mu( \rho_a,\rho_b,z,N) =
\mu( 1-\rho_a,1-\rho_b,{1 \over z},N) 
\label{particle-hole-bis}
\end{equation}

\begin{equation}
\mu( \rho_a,\rho_b,z,N) =
\mu \left( \rho_a,\rho_b,{ \rho_b (1 - \rho_a) \over z \rho_a (1 - \rho_b)},N \right) 
\label{Gallavotti-Cohen-bis}
\end{equation}
{   It is also a fact observed in the perturbation theory to arbitrary
order  (see sections 5 and 6) that, for large
$N$, $\mu$ is proportional   to $1/N$ times a function
of  a single variable $\omega$ defined by}
\begin{equation}
\omega=  {(z-1) (\rho_a z -\rho_b - \rho_a \rho_b(z-1) )
\over z} 
\label{omegadef}
\end{equation}
and the result of our expansion in powers of $\omega$ of section 6 is that
\begin{equation}
\mu = {1 \over N} R(\omega)  + O \left( {1 \over N^2 } \right)
\label{Fmu}
\end{equation}
 where 
\begin{equation}
R(\omega) = \omega  - {\omega^2 \over 3} + { 8 \omega^3 \over 45} - { 4 \omega^4 \over 35} +
O\left( \omega^5 \right)
\label{romegaexp}
\end{equation}
The symmetries (\ref{left-right-bis})-(\ref{Gallavotti-Cohen-bis})
leave  $\omega$ given by (\ref{omegadef})  unchanged so that
 $\mu$ given by (\ref{omegadef}),(\ref{Fmu}) satisfies automatically these symmetries.
From (\ref{romegaexp})  one can easily obtain the large $N$ expression of the  first four cumulants of
$Q(t)$ 
\begin{eqnarray}
&& \lim_{t \to \infty} {\langle Q (t) \rangle \over t} \simeq {1\over N} [
\rho_a - \rho_b]  \label{q1}  \\
&& \lim_{t \to \infty} {\langle Q^2 (t) \rangle_{\rm c} \over t} \simeq {1\over N} \left[
\rho_a + \rho_b - {2 (\rho_a^2 + \rho_a \rho_b + \rho_b^2) \over 3} \right]  
\label{q2} \\
&& \lim_{t \to \infty} {\langle Q^3 (t) \rangle_{\rm c} \over t} \simeq
{1\over N} (\rho_a - \rho_b) \left[ 1  - 2 (\rho_a + \rho_b) +{16 \rho_a^2 + 28
\rho_a \rho_b + 16 \rho_b^2 \over 15} \right]
\label{q3}
\\
&& \lim_{t \to \infty} {\langle Q^4 (t) \rangle_{\rm c} \over t} \simeq
{1\over N}  \left[ \rho_a + \rho_b   - {2 (7 \rho_a^2 + \rho_a \rho_b + 7
\rho_b^2) \over 3} \right.
\label{q4} \\ 
&& \left.
 +{32 \rho_a^3 + 8 \rho_a^2 \rho_b + 8 \rho_a \rho_b^2
+ 32 \rho_b^3  \over 5} - {96 \rho_a^4 + 64 \rho_a^3 \rho_b  - 40
\rho_a^2 \rho_b^2 + 64 \rho_a \rho_b^3 + 96 \rho_b^4 \over 35}  \right]
\nonumber
\end{eqnarray}
One can notice that the $n$th cumulant (at least for $n \leq 4$) 
is a polynomial of degree $n$ in $\rho_a,\rho_b$. We will comment on this at the end of section 5.

\subsection{Two particular cases}
Let us examine these expressions in two particular cases.
First when
\begin{equation}
\rho_a=1 \ \ \  {\rm and}  \ \ \ \rho_b=0
\label{case1}
\end{equation}
one sees that (\ref{q1})-(\ref{q4}) give in the long time limit
\begin{eqnarray}
 {\langle Q(t) \rangle \over t} \to {1 \over N}
\label{q1c1} \\
 {\langle Q(t)^2 \rangle_{\rm c} \over t} \to  {1 \over 3 N}
\label{q2c1} \\
{\langle Q(t)^3 \rangle_{\rm c} \over t}  \to  {1 \over 15 N}
\label{q3c1} \\
{\langle Q(t)^4 \rangle_{\rm c} \over t}  \to  {-1 \over 105 N}
\label{q4c1} 
\end{eqnarray}
These numbers coincide with those found for quantum conductors with many channels in
the metallic regime \cite{LLY} and for quasi-classical conductors
analysed by a Boltzmann-Langevin approach \cite{N}.

Another particular case of interest is when  the two reservoirs are at the same density
$\rho$ 
\begin{equation}
\rho_a = \rho_b = \rho \ .
\label{case2}
\end{equation}
All the odd cumulants vanish and (\ref{q2}),(\ref{q4}) become
\begin{eqnarray}
\lim_{t \to \infty} {\langle Q^2 (t) \rangle_{\rm c} \over t} \simeq
{1\over N}  2 \rho(1-\rho)
\label{q2bis} \\
\lim_{t \to \infty} {\langle Q^4 (t) \rangle_{\rm c} \over t} \simeq
{1\over N}   2 \rho (1-\rho) (1- 2 \rho)^2
\label{q4bis}
\end{eqnarray}

\subsection{Conjecture}
We see in (\ref{q4bis}) that  the fourth cumulant vanishes when $\rho_a=
\rho_b = 1/2$.  We  conjecture that in this particular case, $\rho_a=
\rho_b = 1/2$, all the higher cumulants vanish (i.e. the distribution is
Gaussian)   so that 
 $\mu $ is given in this case by
$$\mu= {1 \over N} {(\log z)^2 \over 4}+  O\left({1 \over N^2 } \right)  \ .$$
This conjecture  (see (\ref{omegadef}),(\ref{Fmu})) fully determines the
function $R(\omega)$
\begin{equation}
R(\omega) = \left[ \log \left (\sqrt{1+ \omega} + \sqrt{\omega} \right)  \right]^2 
\label{Fomega}
\end{equation}
and therefore $\mu$, using (\ref{Fmu}),(\ref{Fomega}) for arbitrary $\rho_a, \rho_b$ and $z$. 

Expression (\ref{Fomega}) needs to be modified when $\omega$ becomes negative.
We will also  conjecture that for $\omega<0$, (\ref{Fomega}) is replaced by its analytic continuation
\begin{equation}
R(\omega) = - \left[ \sin^{-1}  \left(\sqrt{-\omega} \right)  \right]^2 \ .
\label{Fomegabis}
\end{equation}

Looking again at  the first case we analyzed (\ref{case1}), we get 
that not only the first four cumulants (\ref{q1c1})-(\ref{q4c1}) are the
same as those of the distribution first obtained by Lee, Levitov 
and Yakovets \cite{LLY}, but all the higher cumulants are the same
$${\langle Q(t)^5 \rangle_{\rm c} \over t}  \to  {-1 \over 105 N}$$
$${\langle Q(t)^6 \rangle_{\rm c} \over t}  \to  {1 \over 231 N}$$
$${\langle Q(t)^7 \rangle_{\rm c} \over t} \to  {27 \over 5005 N}$$
$${\langle Q(t)^8 \rangle_{\rm c} \over t} \to  {-3 \over 715 N} \ .$$

In the equilibrium case 
 ($\rho_a=\rho_b=\rho $)
too,
this conjecture  determines  all the  cumulants higher than
(\ref{q2bis}),(\ref{q4bis})
$${\langle Q(t)^6 \rangle_{\rm c} \over t}  \to  
  2 { \rho(1-\rho)(2 \rho-1)^2  (1- 16 \rho +16 \rho^2) } $$
$${\langle Q(t)^8 \rangle_{\rm c} \over t}  \to  
2  \rho(1-\rho)(2 \rho-1)^2  (1- 80 \rho +656 \rho^2 -1152 \rho^3 +576
\rho^4)   \ .$$

Our conjecture for the distribution of $Q(t)$ for arbitrary $\rho_a$ and
$\rho_b$ coincides with the distribution found in a multi-channel
quantum picture \cite{GGM} (with a small discrepancy with the
distribution proposed in \cite{Naz}).
 \subsection{The large deviation function}
The knowledge of the large $N$ behaviour of $\mu$ gives some information on the large deviation function $F_N(q)$.
This large deviation function $F_N(q)$ is defined by
\begin{equation}
{\rm Probability}\left( {Q(t) \over t} \simeq {q } \right) \sim
\exp \left[{t } F_N(q) \right]
\label{LDdef}
\end{equation}
or for a more mathematical definition
\begin{equation}
  \lim_{t \to \infty}  {1\over t} \log \left[{\rm Probability}\left(  t q  \leq Q(t) < {tq } + 1 \right)  \right]   =
 F_N(q) 
\label{LDdefmath}
\end{equation}
If we knew $\mu(\alpha,\beta,\delta,\gamma,z,N)$, one would determine
$F_N(q)$ in a parametric form by varying $z$ 
$$q = z {\partial \mu \over \partial z}$$
$$ F_N(q) = \mu - q \log z  \ . $$

\begin{figure}[b!]
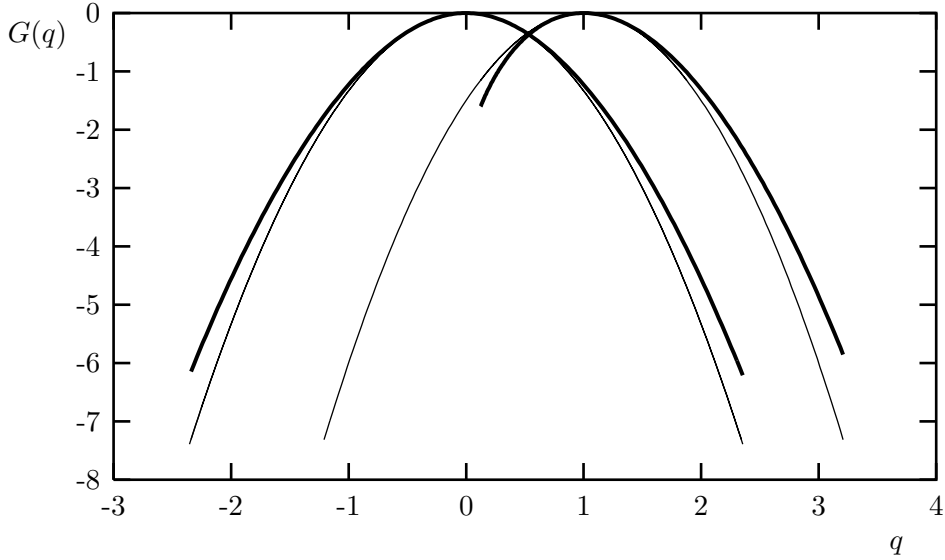

\include{fig5}
\caption{The   rescaled large deviation functions $G(q)$ versus $q$
in the cases $\rho_a=\rho_b= 1/4$ (left thick curve) and  $\rho_a=1$ and
$\rho_b=0$ (right thick curve). The thin lines represent for
comparison the Gaussians with the same two moments.} \label{fig5}
\end{figure}

As here we know only $\mu$ only for large $N$ (see (\ref{Fmu})), we cannot get the full large deviation function $F_N(q)$ for arbitrary $N$  but we can say that in the large $N$ limit,
\begin{equation}
\lim_{N \to \infty} N F_N\left({q \over N} \right) = G(q)
\end{equation}
where the function $G(q)$  can be constructed from $R(\omega)$ in a parametric form 
by varying $z$
\begin{eqnarray}
q = z \left( {d \omega  \over d z} \right) \left( {d R(\omega) \over d
\omega  }\right)
\label{q5}
\\    
G(q)= R(\omega) - q \log z
\label{F5}
\end{eqnarray}
This means that for large $N$  we know $F_N(q)$ only for deviations $q$ of order $1/N$.
Figure 1 shows  $G(q)$ versus $q$ for two choices of $\rho_a$ and $\rho_b$
(the   case  $\rho_a=1$ and $\rho_b=0$ 
and the case $\rho_a=\rho_b=.25$).

\section{The average current and its variance}
In this section we show that the expected value and the variance of the
integrated current $Q(t)$ can be calculated easily by
using the conservation rules.

Let us define $Y_i(t)$ the integrated current between sites $i$ and $i+1$
during the time interval $0,t$
(so $Y_i(t)$ is the total number of particles which have jumped from $i$ to $i+1$  minus the number of particles which have jumped from $i+1$ to $i$ during time $t$).
Similarly let us define $Y_0(t)$ the integrated current from the left reservoir to site $1$ and $Y_N(t)$ the integrated current from site
$N$ to the right reservoir. Note that $Y_0(t)$ and $Q(t)$ have exactly
the same definition and therefore
$$Q(t)=Y_0(t) \ .$$
The conservation of the number of particles implies that
\begin{equation}
Y_{i}(t)= Y_{i-1}(t) +\tau_i(0) - \tau_i(t) \ .
\label{conservation}
\end{equation}
The difference between $Y_i(t)$ and $Y_j(t)$ remains bounded
((\ref{conservation}) implies
that $|Y_{i}(t)- Y_{i-1}(t)  | \leq 2$ and  $|Y_{i}(t)- Y_{j}(t)  | \leq 2 |j-i|$ ).
Therefore in the long time limit the cumulants of $Y_i(t)$ do not depend on $i$.
\begin{equation}
\lim_{t \to\infty}{\log \langle z^{Q(t)}  \rangle \over t }
=
\lim_{t \to\infty}{\log \langle z^{Y_i(t)}  \rangle \over t }
= \lim_{t \to\infty}{\log \langle z^{Y_j(t)} \rangle  \over t }  \ .
\label{conservation-bis}
\end{equation}

The very definition of the dynamics in section 2 means that
during each time interval $dt \ll 1$, 
\begin{eqnarray}
Y_0(t+dt) = & Y_0(t)  \ \ \ & {\rm with \  probability} \ \ \  1 - \alpha (1- \tau_1) dt
- \gamma \tau_1 dt \nonumber \\
 &  Y_0(t) +1   \ \ \ & {\rm with \  probability} \ \ \   \alpha (1- \tau_1) dt \nonumber \\
&   Y_0(t) -1  \ \ \  & {\rm with \  probability} \ \ \   \gamma  \tau_1 dt \nonumber 
\end{eqnarray}
From this evolution  one can deduce the following time evolution
for the moments of $Y_0(t)$:
\begin{equation}
{d   \langle Y_0(t) \rangle \over dt} = \alpha - (\alpha + \gamma)
\langle \tau_1 \rangle 
\label{Y01}
\end{equation}
\begin{equation}
{d   \langle Y_0(t)^2 \rangle \over dt} = 2 \alpha \langle Y_0(t) \rangle  -
2 (\alpha + \gamma)
\langle  Y_0(t) \tau_1 \rangle  + \alpha + (\gamma - \alpha) \langle \tau_1
\rangle 
\label{Y02}
\end{equation}
and more generally
$${d   \langle Y_0(t)^k \rangle \over dt} =  \alpha \langle [ (Y_0(t) +1)^k -
Y_0(t)^k] (1 - \tau_1) \rangle    
 + \gamma 
\langle  [ (Y_0(t) -1)^k  - Y_0(t)^k] \tau_1 \rangle    $$
From (\ref{Y01}),(\ref{Y02})  we obtain
\begin{equation}
{d    \over dt}\langle Y_0(t)^2 \rangle -  \langle Y_0(t) \rangle^2  = -
2 (\alpha + \gamma) [
\langle  Y_0(t) \tau_1 \rangle  - \langle  Y_0(t)  \rangle \langle \tau_1
\rangle ] + \alpha + (\gamma - \alpha) \langle \tau_1 \rangle 
\label{Y02c}
\end{equation}

Similarly starting from the dynamics of the integrated current $Y_i(t) $ through the bond $i,i+1$ or
of  the integrated current $Y_N(t)$ between site $N$ and the right reservoir 
one can  get 
\begin{equation}
{d   \langle Y_i(t) \rangle \over dt} = 
\langle \tau_i \rangle - \langle \tau_{i+1} \rangle 
\label{Yi1}
\end{equation}
\begin{equation}
{d    \over dt}\langle Y_i(t)^2 \rangle -  \langle Y_i(t) \rangle^2  = 2
[\langle Y_i(t) (\tau_i - \tau_{i+1}) \rangle  -
\langle Y_i(t)  \rangle \langle \tau_i - \tau_{i+1} \rangle ] 
 + \langle \tau_i + \tau_{i+1} - 2 \tau_i \tau_{i+1} \rangle
\label{Yi2c}
\end{equation}
and
\begin{equation}
{d   \langle Y_N(t) \rangle \over dt} =
 (\beta + \delta) \langle \tau_N \rangle -  \delta 
\label{YN1}
\end{equation}
\begin{equation}
{d    \over dt}\langle Y_N(t)^2 \rangle -  \langle Y_N(t) \rangle^2  = 2
(\beta + \delta) [ 
\langle Y_N(t) \tau_N  \rangle  -
\langle Y_N(t)  \rangle \langle \tau_N  \rangle ]
 + \delta + (\beta - \delta) \langle \tau_N \rangle 
\label{YN2c}
\end{equation}

\subsection{The current}
If we define the parameters $a,b,\rho_a,\rho_b$ as in \cite{DLS1,DLS2}
and (\ref{rhoa-rhob-def})
 \begin{equation}
a = {1 \over \alpha + \gamma}
 \ \ \ ; \ \ \  
b = {1 \over \beta + \delta} 
\label{abdef}
\end{equation}
$$
\rho_a = {\alpha \over \alpha + \gamma} 
 \ \ \ ; \ \ \  
\rho_b = {\delta \over \beta + \delta} 
$$            
one obtains by combining (\ref{Y01}),(\ref{Yi1}),(\ref{YN1}) that
$$ a {d \langle Y_0 \rangle \over dt}
+ b {d \langle Y_N \rangle \over dt}
+ \sum_{i=1}^{N-1} {d \langle Y_i \rangle \over dt}= \rho_a - \rho_b \ . $$
We know (\ref{conservation},\ref{conservation-bis}) that in the
steady state  $d \langle Y_i \rangle  /dt $ does not depend on $i$.
Therefore, one  obtains
that way the steady state current 
\begin{equation}
  {d \langle Q \rangle \over dt} =
  {d \langle Y_i \rangle \over dt} = {\rho_a - \rho_b \over N + a +b -1}
\label{steady-state-current}
\end{equation}
which gives (\ref{q1}) for large $N$.

\subsection{The variance}
Similarly  adding  (\ref{Y02c}),(\ref{Yi2c}),(\ref{YN2c}) and
using the fact
that in the steady state $ {d   \langle Y_i^2 \rangle -  \langle Y_i
\rangle^2 \over dt}$ does not depend on $i$ one gets
\begin{eqnarray}
(a + b + N-1) &  {d   \langle Y^2 \rangle -  \langle Y \rangle^2 \over dt} =
  2 \sum_{i=1}^N  \langle Y_i \tau_i \rangle - \langle Y_i \rangle
\langle \tau_i \rangle - \langle Y_{i-1} \tau_i \rangle + \langle Y_{i-1}
\rangle \langle \tau_i \rangle 
\nonumber \\
&
+ \rho_a + \rho_b
   - 2 \rho_a \langle \tau_1 \rangle  
  - 2 \rho_b \langle \tau_N \rangle   +  2 \sum_{i=1}^{N} \langle \tau_i
\rangle -  2 \sum_{i=1}^{N-1} \langle \tau_i \tau_{i+1}
\rangle
\nonumber
\\
\label{variance1}
\end{eqnarray}
and using (\ref{conservation}) one obtains (using that 
 $\langle \tau_i(0) \tau_i (t) \rangle \to \langle \tau_i(0)\rangle
\langle  \tau_i
\rangle $ in the long time limit)
\begin{equation}
(a + b + N-1) {d   \langle Y^2 \rangle -  \langle Y \rangle^2 \over dt} =
 \rho_a + \rho_b - 2 \rho_a \langle \tau_1 \rangle - 2 \rho_b \langle
\tau_N \rangle 
+   2 \sum_{i=1}^N  \langle  \tau_i \rangle^2 - 
2 \sum_{i=1}^{n-1} \langle \tau_i \tau_{i+1} \rangle 
\label{variance2}
\end{equation}
All the steady state correlations can be
calculated exactly \cite{HS1,DLS2}, in particular
$$\langle \tau_i \rangle = \rho_b +{N-i+b \over N+a+b-1} (\rho_a-
\rho_b) = {\rho_a (N+b-i) + \rho_b (i-1+a) \over N+a+b-1} $$
 and for $i < j$
$$\langle \tau_i \tau_j \rangle - \langle \tau_i \rangle \langle \tau_j
\rangle =  - (\rho_b-\rho_a)^2 {(a+ i -1) (b+ N - j) \over (N+a+b-1)^2
(N+a+b-2) } $$
 so that (\ref{variance2}) becomes
\begin{eqnarray}
 {d  [ \langle Q^2 \rangle -  \langle Q \rangle^2 ] \over dt} =
 {d  [ \langle Y_i^2 \rangle -  \langle Y_i \rangle^2 ] \over dt} ={1 \over
N_1}( \rho_a +
\rho_b - 2 \rho_a \rho_b) 
\nonumber \\
 + {a (a-1) (2 a -1) + b
(b-1) (2b-1) - N_1 (N_1-1)(2 N_1-1) \over 3 N_1 ^3 (N_1-1)}
(\rho_a - \rho_b)^2 
\nonumber \\
\label{variance-final}
\end{eqnarray}
where $ N_1 = N+a+b -1$. In the large $N$ limit, one obtains (\ref{q2}).

\section{A hierarchy  of equations for the correlation functions }
In the long time limit, the vector  $ {\cal P}_t({\cal C},z) $ in
(\ref{PCz}) becomes an eigenvector of the matrix $M_z$ defined in
(\ref{Mz}).
$$ {\cal P}_t({\cal C},z) \sim   e^{\mu  t} \  \psi_\mu({\cal C})
$$ where  $\psi_\mu({\cal C})$ satisfies
\begin{equation}
\mu \psi_\mu({\cal C}) =  \sum_{\cal C'} M_z({\cal C},{\cal C'})
\psi_\mu({\cal C'}) 
\label{psi}
\end{equation}

From (\ref{psi}), one can build a hierarchy of equations which, as  we
shall see it 
 in the next section 5, can be truncated either when one
expands in powers of $z-1$ to obtain the first cumulants or when the
densities $\rho_a$ and $\rho_b$ in the reservoirs are small.

Let us define the following correlation functions: \\
 for $1 \leq i \leq N$
\begin{equation}
T_i = \sum_{\cal C} \psi_\mu({\cal C}) \tau_i({\cal C})
\label{Ti}
\end{equation}
for $1 \leq i < j \leq N$
\begin{equation}
U_{i,j} = \sum_{\cal C} \psi_\mu({\cal C}) \tau_i({\cal C}) \tau_j({\cal
C})  
\label{Uij}
\end{equation}
for $1 \leq i < j < k \leq N$
\begin{equation}
V_{i,j,k} = \sum_{\cal C} \psi_\mu({\cal C}) \tau_i({\cal C}) \tau_j({\cal
C})  \tau_k({\cal C}) 
\label{Vijk}
\end{equation}
and so on, with the convention
that
\begin{equation}
 \sum_{\cal C} \psi_\mu({\cal C})  =1 
\label{norma}
\end{equation}

Inserting these definitions into (\ref{psi}), one obtains a hierarchy of
equations for the one-point functions $T_i$, the two point functions
$U_{i,j}$, and so on.
By summing (\ref{psi}) over all ${\cal C}$, one obtains
\begin{equation}
\mu = \alpha ( z -1) +  \left( \gamma {1 \over z}   - \alpha z + \alpha -
\gamma\right)  T_1
\label{mu1}
\end{equation}
By multiplying (\ref{psi}) by $\tau_i({\cal C})$ and summing over $ {\cal
C}$, one gets
\begin{equation}
\mu  T_i = \alpha ( z -1)T_i  + 
 \left( \gamma {1 \over z} - \alpha z + \alpha - \gamma \right) U_{1,i}
 + T_{i-1} + T_{i+1} - 2 T_i 
\label{Ti1}
\end{equation}
At the two boundaries (\ref{Ti1}) is modified
\begin{equation}
\mu T_1 = \alpha  z  - (  \alpha z  +  \gamma) T_1 + T_2 - T_1 
\label{T11}
\end{equation}
\begin{equation}
\mu  T_N = \alpha ( z -1)T_N  +
\left( \gamma {1 \over z} - \alpha z + \alpha - \gamma \right) U_{1,N} 
 + T_{N-1} - (1+\beta+\delta ) T_N + \delta 
\label{TN1}
\end{equation}
In fact (\ref{T11}) and (\ref{TN1}) (which are the boundary versions of
(\ref{Ti1}))
reduce to (\ref{Ti1}) provided that  we require that $T_0$,
$T_{N+1}$ and $U_{1,1}$
(for non-physical values of the parameters) 
satisfy
\begin{equation}
 \alpha (z-1) T_1 + \left( \gamma {1 \over z} - \alpha z + \alpha -
\gamma \right)
 U_{1,1}  + T_0 - T_1  = \alpha
z - (\alpha z + \gamma) T_1
\label{boundary1}
\end{equation}
\begin{equation}
\delta - (\beta + \delta) T_N = T_{N+1} - T_N 
\label{boundary2}
\end{equation}
Similarly by multiplying    by $\tau_i({\cal C}) \tau_j({\cal C})$ one gets
\begin{eqnarray}
 \mu U_{i,j} = \alpha  (z-1) U_{i,j}  + \left( \gamma {1 \over z} - \alpha z +
\alpha - \gamma \right) V_{1,i,j}
 + U_{i-1,j} + U_{i+1,j}  && \nonumber \\ + U_{i,j-1} +
U_{i,j+1} - 4 U_{i,j}  && 
\label{Uij1}
\end{eqnarray}
the boundary conditions  and the case $j=i+1$ being automatically satisfied provided that the
extensions of 
$U_{0,i}, U_{i,i}, U_{i,N+1}, V_{1,1,i}$ to non-physical values satisfy
\begin{equation}
\alpha  (z-1) U_{1,i}  + \left( \gamma {1 \over z} - \alpha z + \alpha -
\gamma \right) V_{1,1,i} 
 + U_{0,i} - U_{1,i} = \alpha z T_i -
 (\alpha z+ \gamma)U_{1,i}
\label{boundary3}
\end{equation}
\begin{equation}
\delta T_i - (\beta + \delta) U_{i,N} = U_{i,N+1} - U_{i,N}
\label{boundary5}
\end{equation}
\begin{equation}
U_{i,i}+U_{i+1,i+1} = 2 U_{i,i+1}
\label{boundary4}
\end{equation}
(Note that   definitions  such as (\ref{Uij})
 do not tell us what $U_{i,i}$  is as $U_{i,j}$ is only defined for
$j> i$. In general the values one has to choose for non-physical
values of the parameters in order to satisfy the boundary conditions
 (\ref{boundary3})-(\ref{boundary4}) are different from what one could obtain by simply putting $j=i$ in the
definition:
in particular $U_{i,i} \neq T_i$. In this whole paper the $U_{i,j}$ we
calculate are always polynomials in $i$ and $j$ and the unphysical values
such as $U_{i,i}$ are simply  obtained by taking $j=i$ in the polynomial
$U_{i,j}$).

All the relations for the higher correlation functions can be generated
along the same steps.
One  way of writing all these relations is to introduce the
generating function 
  $$\Phi(a_1,...a_L;z) = \left\langle z^{Q(t)}  \exp \left[\sum_i a_i
\tau_i(t) \right]
\right\rangle $$
(where  $Q(t)$, as above,  is the total number of particles transferred from the
left
  reservoir to site $1$ during time $t$). 
For large $t$ one expects that $$
\Phi(a_1,...a_L) \sim e^{\mu \; t}  \ \phi(a_1,...a_L) $$
where $\phi$ satisfies
\begin{eqnarray}
\mu \phi = \left[ \sum_{i=1}^{L-1}
( e^{a_{i+1}-a_i} -1) \left(
{\partial \over \partial a_i} - {\partial^2 \over \partial a_i  \partial
a_{i+1}} \right) \right.
\nonumber \\
+ ( e^{a_{i}-a_{i+1}} -1) \left(
{\partial \over \partial a_{i+1}} - {\partial^2 \over \partial a_i
\partial
a_{i+1}} \right)   \nonumber \\
+ \alpha \left( z e^{a_1} - 1 \right)\left( 1  - {\partial \over
\partial a_1 } \right)
+ \gamma \left(  {e^{-a_1} \over z}  - 1 \right) {\partial \over \partial
a_1 }
\nonumber \\
\left. + \delta \left(  e^{a_L} - 1 \right)\left( 1  - {\partial \over \partial
a_L } \right)
+ \beta \left( e^{-a_L}   - 1 \right) {\partial \over \partial a_L }
 \right] \phi
\label{phi}
\end{eqnarray}
Expanding  (\ref{phi}) in powers of the $a_i$ allows one to recover all the
above relations between the correlation functions
(\ref{mu1}),(\ref{Ti1}) ....,
and to generate the equations satisfied by the higher correlations.
The first levels of the hierarchy are summarized in  Appendix \ref{hierarchy}.

\section{The low density expansion}
When the densities $\rho_a$ and $\rho_b$ of the reservoirs are small 
the $n$-point function is  of order $n$ to leading order in $\rho_a$ and
$\rho_b$. 
To calculate $\mu$ to order $n$ in $\rho_a$ and $\rho_b$,
one can truncate the hierarchy by neglecting all the $m$-point correlation
functions for $m>n$.

A priori the truncated hierarchy remains a problem hard to solve.
However we noticed that {\it the solutions $T_i$, $U_{i,j}$ ... of the truncated
hierarchy are always polynomials in the coordinates $i,j..$}
(see the Appendix \ref{ruin} on the analogy with a multi-particle ruin problem).
For example if one tries to expand to order $3$ in $\rho_a$ and $\rho_b$
(or in $\alpha$ and $\gamma$) one finds that $T_i$ is a polynomial of degree $5$ in $i$,
$U_{i,j}$ of degree 4 in $i,j$ and $V_{i,j,k}$ a polynomial of degree 3
in $i,j,k$ (in fact $V_{i,j,k}$ is linear in each of the three
coordinates).
So to solve the truncated hierarchy, we introduced arbitrary parameters
(the coefficients of all the polynomials in $i$,  \ $i,j$, \ $i,j,k $ ...)
and the equations of the hierarchy give us a finite set of linear
equations to solve for these parameters.

We used Mathematica to solve these linear equations. The expressions
become quickly complicated for general $a,b,\rho_a,\rho_b$.
The general expression  (\ref{mu3}) of $\mu$ at order $3$ in $\rho_a$ and $\rho_b$ is
given in the Appendix \ref{hierarchy}.
We give here the result  obtained that way for $\mu$ to order $3$ in
$\rho_a,\rho_b$ when $a=b=1$

\begin{eqnarray}
&& \mu={(z-1) (\rho_a z  - \rho_b) \over z (N+1)} 
 \\
&& - {(z-1)^2 ( \rho_b^2 + 4 \rho_a \rho_b z + \rho_a^2 z^2 + 2 N (\rho_b^2
+ \rho_a \rho_b z + \rho_a^2 z^2 )) \over 6 z^2 (N+1)^2 }
\nonumber \\
&& + {(z-1)^3 ( 2 N+1) (\rho_a z - \rho_b) ( 3 \rho_b^2 + 9 \rho_a \rho_b z
+ 3 \rho_a^2 z^2 +  N ( 4 \rho_b^2
+  7 \rho_a \rho_b z +  4 \rho_a^2 z^2 )) \over 45 z^3 (N+1)^3 }
\nonumber
\end{eqnarray}

For large $N$ the expression (\ref{mu3})  of $\mu$ gets much simpler: to
leading order in $N$, the results do not depend anymore on the two
parameters $a$ and $b$ and one gets 
\begin{eqnarray}
\mu = {1 \over N} \left[ {(\rho_a z - \rho_b ) (z-1) \over z}
- { (\rho_a^2 z^2 + \rho_a \rho_b z + \rho_b^2 )  (z-1)^2 \over 3 z^2}
\right.
\nonumber \\ \left.
+ {2 ( \rho_a z - \rho_b) ( 4 \rho_a^2 z^2 + 7 \rho_a \rho_b z +
\rho_b^2) ( z-1)^3 \over 45 z^3}  + O \left( \rho^4 \right) \right]
\label{mu3scaling}
\end{eqnarray}
So in this large $N$  regime, $\mu$ is proportional to $1/N$ and is a function
of three parameters $\rho_a,\rho_b$ and $z$.
In fact, if one uses the parameter 
$ \omega = (z-1) (\rho_a z -  \rho_b - \rho_a \rho_b (z-1)) / z $
 defined in (\ref{omegadef}),
one can easily check that
(\ref{mu3scaling}) can be rewritten as
\begin{equation}
\label{mu3scalingbis}
\mu= {1 \over N} \left( \omega - {2 \over 3} \omega^2 + {8 \over 45} \omega^3 + O(\omega^4)
\right)
\end{equation}
Up to the factor $1/N$, $\mu$ depends on the  single parameter $\omega$,
defined by (\ref{omegadef}), (at least to order 3 in $\omega$).
The expansion of $\mu$ in powers of $\rho_a$ and $\rho_b$
to third order determines exactly the first three cumulants and more
generally the expansion of $\mu$ to order $n$ would give the exact
expression of the first $n$ cumulants.
This can be understood by noticing the similarity between an expansion of
$\mu$ in powers of $z-1$ and  an expansion of $\mu$ in powers of $\rho_a$
and $\rho_b$. In both cases, the hierarchy can be truncated and 
one  can neglect all the correlations higher than the $n$-point
function if one wishes to obtain $\mu$ at order $n$. 
This is the reason why the exact expression of the  $n$th cumulant
is a polynomial of degree $n$ in $\rho_a$ and $\rho_b$ as noticed
at the end of section 2.3.
\section{Continuous limit}

The expressions of the $T_i$'s, $U_{i,j}$'s , $V_{i,j,k}$'s  we have obtained by
Mathematica to solve the hierarchy 
in powers of  $\rho_a$ and $\rho_b$ are rather complicated. However they take a somewhat simpler form in the large $N$
limit. If one considers the connected correlation functions $u_{i,j},
v_{i,j,k}$ ...
defined by (\ref{uijdef}), (\ref{vijkdef}), 
 their expressions  become functions of the
continuous variables:
\begin{equation}
x_1 = {i \over N} \ \ \ , \ \ \ 
x_2 = {j \over N} \ \ \ , \ \ \ 
x_3 = {k \over N} \  .
\label{x1x2x3def}
\end{equation}
To leading order in $1/N$ and to third order in powers of $\rho_a$ and
$\rho_b$, one obtains that way:
\begin{eqnarray}
T_i \simeq  \rho_a z \left[ r x_1 + (1 - x_1) \left( 1 + s(-1 + (1 - r)^2
x_1 /3  - (1-r)^2
x_1^2 /6)  \right. \right. 
  \nonumber \\  
 +  s^2
(1 +(-23 + 24 r - 9 r^2 + 8 r^3) x_1 / 45 + (29 - 42 r + 27 r^2 - 14 r^3)
x_1^2 /90 
  \nonumber \\  \left. \left.
+ (r-1)^3 x_1^3 / 15 - (r-1)^3 x_1^4 /60 ) \right) \right]
\end{eqnarray}
\begin{eqnarray}
u_{i,j} \simeq {1 \over N}  \rho_a^2 z^2  \left[ x_1 (1-x_2) \left(-(1-r)^2 + s
( 2 (4 - 6 r + 3 r^2 - r^3) /3 +
\right. \right.  \nonumber \\\left. \left. (r-1)^3 x_1 -
(r-1)^3 x_1^2 /3 + 2 (r-1)^3 x_2 /3 - (r-1)^3 x_2^2 /3) \right) \right]
\end{eqnarray}
\begin{eqnarray}
v_{i,j,k} \simeq {1 \over N^2}  \rho_a^3 z^3  \left[ -2 x_1 (1 - 2 x_2)
(1-x_3) \right]
\end{eqnarray}
where
the parameters $r$ and $s$ are defined by
\begin{equation}
r= {\rho_b \over \rho_a z} 
\label{rdef}
\end{equation}
\begin{equation}
s=  \rho_a (z-1)  \ .
\label{sdef}
\end{equation}

By examining the hierarchy (see Appendix \ref{hierarchy}) for the connected functions
and by assuming that the structure obtained up to  third order
persists to higher orders  one expects   that to leading order in $N$
\begin{equation}
 T_i  \simeq \rho_a z f(x_1) 
\label{scal1}
\end{equation}
\begin{equation}
 u_{i,j}  \simeq {\rho_a^2  z^2 \over N}  g(x_1,x_2) 
\label{scal2}
\end{equation}
\begin{equation}
 v_{i,j,k}  \simeq {\rho_a^3  z^3 \over N^2}  h(x_1,x_2,x_3)  \ .
\label{scal3}
\end{equation}
With this scaling and as $\mu$ is of order $1/N$, one can neglect the right hand side of
(\ref{T13})-(\ref{V1ij3}). One can even show that
$g(0,x_2)=h(0,x_2,x_3)=i(0,x_2,x_3,x_4)=0$ so that 
 (see (\ref{epsilondef}))
\begin{equation}
T_1  \simeq  { \alpha z \over \alpha z + \gamma}
\label{rel1}
\end{equation}
\begin{equation}
\epsilon \  u_{1,i} \simeq \left(1 - {1 \over z} \right) (1 + \rho_a(z-1)) (u_{1,i} - u_{0,i}) 
\label{rel2}
\end{equation}
\begin{equation}
\epsilon \  z_{1,i,j} \simeq \left(1 - {1 \over z} \right) (1 + \rho_a(z-1)) (z_{1,i,j} - z_{0,i,j}) 
\label{rel3}
\end{equation}
and  with these simplifications 
  the hierarchy (\ref{mup3})-(\ref{Vijj3})  in the large $N$ regime
becomes:
\\
\underline{the equation for $\mu$} 
\begin{equation}
\mu= s (1+s) f'(0) 
\label{hie1}
\end{equation}
\underline{the bulk equations} (\ref{Ti3}-\ref{Vijk3}) 
\begin{equation}
 s(1+s) {d \over d x_1} g(0,x_2) = { {d^2 \over d
x_1^2}} f(x_2)
\label{hie2}
\end{equation}
\begin{equation}
 s(1+s) {d \over d x_1} h(0,x_2,x_3) = \left(
{d^2 \over dx_1^2} + {d^2 \over dx_2^2} \right) g(x_2,x_3) 
\label{hie3}
\end{equation}
\begin{equation}
 s(1+s) {d \over d x_1} i(0,x_2,x_3,x_4) = \left(
{d^2 \over dx_1^2} + {d^2 \over dx_2^2} + {d^2 \over dx_3^2}\right)
h(x_2,x_3,x_4) 
\label{hie4}
\end{equation}
\underline{the left and right boundary  equations}
\begin{equation}
f(0)= {1 \over 1+s}  \ \ \ ; \ \ \ 
f(1)= r 
\label{hie5}
\end{equation}
\begin{equation}
g(0,x)=g(x,1)=0 
\label{hie6}
\end{equation}
\begin{equation}
h(0,x,y)=h(x,y,1)=0 
\label{hie7}
\end{equation}
\underline{the  equations for adjacent particles}
\begin{equation}
 \left({d \over d x_1} - {d\over dx_2} \right) g(x,x) = - \left( {df(x)
\over dx_1 }\right)^2 
\label{hie8}
\end{equation}
\begin{equation}
 \left({d \over d x_1} - {d\over dx_2} \right) h(x,x,y) = -
2  {d f(x) \over dx_1}  {d g(x,y) \over dx_1}  
\label{hie9}
\end{equation}
\begin{equation}
 \left({d \over d x_2} - {d\over dx_3} \right) h(x,y,y) = -
2  {d f(y) \over dx_1}  {d g(x,y) \over dx_2}  
\label{hie10}
\end{equation}
\begin{equation}
 \left({d \over d x_1} - {d\over dx_2} \right) i(x,x,y,z) = -
2  {d f(x) \over dx_1}  {d h(x,y,z) \over dx_1}
-2  {d g(x,y) \over dx_1}  {d g(x,z) \over dx_1} 
\label{hie11}
\end{equation}
\begin{equation}
 \left({d \over d x_2} - {d\over dx_3} \right) i(x,y,y,z) = -
2  {d f(y) \over dx_1}  {d h(x,y,z) \over dx_2}
-2  {d g(x,y) \over dx_2}  {d g(y,z) \over dx_1}
\label{hie12}
\end{equation}
\begin{equation}
 \left({d \over d x_3} - {d\over dx_4} \right) i(x,y,z,z) = -
2  {d f(z) \over dx_1}  {d h(x,y,z) \over dx_3}
-2  {d g(x,z) \over dx_2}  {d g(y,z) \over dx_2} 
\label{hie13}
\end{equation}

One can then solve this hierarchy, up to an arbitrary order in $s$. To
find $\mu$ at order $n$ in $s$, one needs to know $f$ at order $n-1$, $g$
at order $n-2$ and so on.
When we calculated $\mu$ at order $4$ in $s$, we obtained
{\small
\begin{eqnarray}
f(x_1)= r +\left( r - {1 \over s+1} \right)(x_1 -1) \left[  1 - \omega {x_1  
 (x_1-2) \over 6}    
+
 \omega^2 {x_1 (3 x_1^3 - 12 x_1^2 + 28 x_1 - 32) \over  180}  \right. 
\nonumber \\
\left.  - \omega^3 {x_1  ( 5 x_1^5 - 30 x_1^4 +138  x_1^3 - 352 x_1^2 + 600
   x_1 - 576) \over  5040}    \right]
\end{eqnarray}
\begin{eqnarray}
g(x_1,x_2)= \left( r - {1 \over s+1} \right)^2  x_1 (x_2-1) \left[  1 -
\omega { 2 - 3 x_1 + x_1^2 - 2 x_2 + x_2^2 
 \over 3}    \right. 
\nonumber \\ 
\left. +
\omega^2 \left( {x_1^4 \over  24} - {x_1^3\over  4} + { x_1^2 ( 26 - 10 x_2 + 5
x_2^2) \over 36} - {x_1 ( 3 - 2 x_2 + x_2^2) \over 3} + { 56 - 80 x_2 + 60
x_2^2 - 20 x_2^3 + 5
x_2^4 \over 120} \right)
\right]
\nonumber \\
\end{eqnarray}
\begin{eqnarray}
h(x_1,x_2,x_3)= \left( r - {1 \over s+1} \right)^3  x_1 (x_3-1) \left[
4 x_2 - 2 +
    \right. 
\nonumber \\ 
\left. -
 \omega { 5 x_2^3 - 15 x_2^2 +  5 x_2 (  x_1^2 - 3 x_1+   x_3^2 - 2 x_3 +
4)  - 3 x_1^2 + 9 x_1 - 2 x_3^2 + 4 x_3 - 6
\over 3}  
\right]
\end{eqnarray}
\begin{eqnarray}
i(x_1,x_2,x_3,x_4)= \left( r - {1 \over s+1} \right)^4   2 x_1 (x_4-1) 
( 15 x_2 x_3 - 10 x_2 - 5 x_3 + 3)
\end{eqnarray}
}
and \begin{eqnarray}
\mu = {1 \over N} \left[\omega -{\omega^2 \over 3} + {8  \omega^3 \over
45} - {4 \omega^4 \over 35} + O\left( \omega^5 \right) \right]
\end{eqnarray}
where we give the expressions of $f,g,h,i$ and $\mu$ (except for  a simple factor
$r - {1 \over 1 + s}$) in powers of $\omega$ defined by (\ref{omegadef}) instead
of $s$.

 In principle all the expressions should  depend on the two parameters $s$ and $r$,  but we observe that they only depend on
the single
 parameter $\omega$.
This can be understood by noticing that if $f,g,h...$ solve the hierarchy
(\ref{hie1})-(\ref{hie13}) 
for a certain choice of  $r,s$, then $A f + B$, $A^2 g$, $A^3 h$ ...
solve the same hierarchy for $r',s'$  with the same value of $\mu$ if $r'$ and $s'$ satisfy 
 $${1 \over 1 + s'} = A
{1 \over 1 + s} + B $$
$$ r' = A r + B $$
$$A s' (1+ s') = s (1+s) $$
These three relations are compatible only when $$s' - r' s' - r' s'^2 =
s - r  s - r s^2 $$
 so that when $\omega= s - r s - r s^2 $ remains unchanged,
one can easily transform  the solution of the hierarchy leaving  $\mu$
unchanged.
\section{Conclusion}
In this paper we have obtained the first four cumulants
(\ref{q1})-(\ref{q4}) of the integrated
current for the symmetric simple exclusion process with open boundaries.
To our surprise, the generating function of the integrated current
(\ref{mudef}),(\ref{Fmu}) depends
on the densities of the reservoirs $\rho_a$ and $\rho_b$ and on the fugacity $z$,  the 
parameter conjugated to the integrated current, through a single
parameter $\omega$
defined in (\ref{omegadef}). It would be interesting to understand why this is so through a simple physical argument.

When $\rho_a=\rho_b=1/2$,  the fourth cumulant vanishes and we have
conjectured that in this particular case, the distribution of the
integrated current  $Q(t)$ is Gaussian (in the range ${Q(t) \over t} \sim {1
\over N} $). Based on this conjecture, we can predict 
(\ref{Fomega}),(\ref{Fomegabis}) the large deviation function of the
current for arbitrary choices of $\rho_a,\rho_b$. For $\rho_a=1$ and
$\rho_b=0$, the distribution of the integrated current we obtained is
identical to the one known for one dimensional  quantum conductors in
their metallic regime \cite{LLY,BlB}.

The similarity between these results is striking if we consider the
drastic differences in the corresponding formalisms.
In the quantum treatment of a diffusive conductor, the statistics
of the time integrated current appears as the result of a convolution of
a large number of independent binomial laws, one for each conduction
channel \cite{LLY}.
In the limit of a large number of such channels (i.e. when the transverse
dimension of the conductor is much larger than the Fermi wave length)
the result of this convolution is governed by the universal distribution
of eigenvalues of the transmission matrix for a single particle in the
presence of quenched disorder. The exclusion effects induced by the Pauli
principle
only appear in the selection of the energy window in which single
particle states contribute to the current.
By contrast, the classical model considered here has no transverse degree
of freedom, and the exclusion  constraint plays a crucial role.
To our knowledge, a complete understanding of the connection between the
two models is still lacking.
We simply conjecture that an intermediate description in terms of a
Boltzmann equation with additional noise terms,  as developed  for
instance in \cite{N0,N} for the quantum diffusive case, may
help to bridge the gap between the  two  classes of systems.

The first open question left at the end of the present paper is whether 
one could prove  or disprove our conjecture for $\mu$ in section 2.5.
It would also be interesting to see the degree of universality of the
results obtained here, i.e.  how much they depend on the precise
definition of the model.
In particular it would be  nice to see whether  a more macroscopic
approach 
could be used to calculate the fluctuations of the current
\cite{BDGJL,BDGJL2}.
Another open question would be   to know how our results would be
modified by an asymmetry \cite{DEHP,Sas,BECE} in the bulk, in particular in the case
of a weak asymmetry \cite{ED}.
\newpage
\appendix

\section{ The Gallavotti-Cohen relation \label{GCRa}}
In this appendix we rederive, following Lebowitz and Spohn \cite{LS},
the Gallavotti-Cohen relation for a system with stochastic dynamics in contact
with several reservoirs of particles.

Let us consider an irreducible  Markov process for a system with a finite number of internal configurations
$\cal C$. We assume that this system is in contact with a reservoir A (or several reservoirs A,B,C)
and that during each infinitesimal time interval $dt$, there is a probability
$W_q({\cal C'},{\cal C }) dt$ of a jump from ${\cal C }$ to ${\cal C'}$
with  $q$ particles
transferred from  reservoir A  to the system during this jump.
As the system is in general in contact with other reservoirs,  these particles might  later on be transferred
to other reservoirs, so that $W_0({\cal C'},{\cal C }) $ allows jumps
where the number of particles in the system is not conserved.

Imagine that the system is in equilibrium with  reservoir A, that is  the jumping rates $W_q({\cal C'},{\cal C })$
satisfy the detailed balance condition
\begin{equation}
W_q({\cal C'},{\cal C }) P_{\rm eq}({\cal C })
= W_{-q}({\cal C},{\cal C' }) P_{\rm eq}({\cal C' })
\label{detailedbalance}
\end{equation}
where
 $P_{\rm eq}({\cal C })$ is the steady state probability of the Markov
process.

Clearly the detailed balance condition  (\ref{detailedbalance}) implies that the average current of particles vanishes
and that the probability of seeing any given jump is equal to the
probability of its time reversal as it should for a system at equilibrium.

Now let us modify the dynamics by introducing a field $E$ which enhances the injection of particles into the system
so that $W_q({\cal C'},{\cal C }) $  is replaced by
\begin{equation}
e^{E q } W_q({\cal C'},{\cal C }) 
\label{non-zero-field}
\end{equation}
This field $E$  produces a current which of course
fluctuates due to the stochastic nature of the Markov process.

Let us denote by $Q(t)$ the total number of particles transferred from
reservoir A to the system during time $t$
and $R_t({\cal C },Q)$ the  probability of $Q(t)$, given  that the system is in  configuration 
$\cal C$ at time  $t$.
The evolution of  $R_t({\cal C },Q)$ is clearly
$${d \over dt}  R_t({\cal C },Q) = \sum_q  
 \sum_{\cal C' } e^{E q }
\left[
W_q({\cal C},{\cal C' })    R_t({\cal C' },Q-q) -
W_q({\cal C'},{\cal C })    R_t({\cal C },Q) \right]
$$
If one introduces the generating functions $r_t({\cal C},\lambda)$ defined by
$$r_t({\cal C},\lambda)= \sum_Q e^{\lambda Q }  R_t({\cal C },Q)$$
they evolve according to
$${d \over dt}  r_t({\cal C },\lambda) = \sum_q  
 \sum_{\cal C' } 
\left\{
e^{(E +\lambda ) q}
W_q({\cal C},{\cal C' })    r_t({\cal C' },\lambda) -
e^{E  q}
W_q({\cal C'},{\cal C })    r_t({\cal C },\lambda) \right\}
$$
This implies  that for large $t$, 
$$\langle e^{\lambda Q(t)} \rangle \sim e^{\mu(\lambda,E) t } $$
where $\mu(\lambda,E)$ is the largest eigenvalue of the matrix $M_{\lambda,E}$
\begin{equation}
 M_{\lambda,E} = \sum_q
e^{(E +\lambda ) q} \ 
W_q({\cal C},{\cal C' })    -
\delta({\cal C},{\cal C' }) 
 \sum_q \sum_{\cal C'' } e^{E  q}  \  W_q({\cal C''},{\cal C })   
\label{matrix}
\end{equation}
where $\delta({\cal C},{\cal C' })=1$ if ${\cal C}={\cal C' }$ and $0$ if
${\cal C} \neq {\cal C' }$.
Therefore  to obtain $\mu(\lambda,E)$, one has to find either the right eigenvector $\psi_R({\cal C})$ of this matrix
which satisfies
\begin{equation}
\mu(\lambda,E)  \psi_R({\cal C }) = 
\sum_q \sum_{\cal C' }
e^{(E +\lambda ) q} \ 
W_q({\cal C},{\cal C' })   \psi_R({\cal C' }) -
\sum_q \sum_{\cal C' }
e^{E  q} \ 
W_q({\cal C'},{\cal C })  \psi_R({\cal C }) 
\label{right}
\end{equation}
or its left eigenvector
$\psi_L({\cal C})$
\begin{equation}
\mu(\lambda,E)  \psi_L({\cal C }) = \sum_q
 \sum_{\cal C' }
e^{(E +\lambda ) q} \ 
W_q({\cal C'},{\cal C })   \psi_L({\cal C' }) -
\sum_q \sum_{\cal C' }
e^{E  q} \ 
W_q({\cal C'},{\cal C })  \psi_L({\cal C })
\label{left}
\end{equation}

Now if we use the detailed balance condition (\ref{detailedbalance}) for
the first term in the r.h.s. of (\ref{left}),
we get
\begin{equation}
\mu(\lambda,E)  \psi_L({\cal C }) = \sum_q
 \sum_{\cal C' }
e^{(E +\lambda ) q}
W_{-q}({\cal C},{\cal C' })
 {P_{\rm eq}({\cal C' }) \over 
 P_{\rm eq}({\cal C })}
   \psi_L({\cal C' }) -
\sum_q \sum_{\cal C' }
e^{E  q}
W_q({\cal C'},{\cal C })  \psi_L({\cal C })
\label{leftm}
\end{equation}
This shows that ${\psi_L({\cal C })  P_{\rm eq}({\cal C })}$ is the {\it right} eigenvector
of the matrix $M_{- \lambda - 2 E,E}$ defined in (\ref{matrix}).

So the matrices $M_{\lambda,E}$ and $M_{-\lambda - 2 E,E}$ have exactly the same eigenvalues.
In particular this shows that
\begin{equation}
\mu(\lambda,E) = \mu(-\lambda - 2 E,E)
\label{GC}
\end{equation}
which is the Gallavotti-Cohen relation.

In the symmetric exclusion process, as described in section 2.1, we know
\cite{DLS2}
 that detailed balance is satisfied whenever
\begin{equation}
{ \alpha \over \alpha + \gamma}={ \delta \over \beta + \delta} 
\label{db}
\end{equation}
If we fix $\beta$ and $\delta$ and  vary 
$\alpha$ and $\gamma$, the detailed balance condition (\ref{db}) is no longer verified.
However, one can always think of the variation of $\alpha$ and $\gamma$ as the effect of an external field $E$ 
trying to enhance the number of particles transferred from the left reservoir to site $1$.
If one writes
$$ \alpha = \alpha' e^E \  \ \ \ {\rm and } \ \ \   \gamma = \gamma' e^{-E}$$
with $$\alpha'= {\delta \over \beta} \gamma'=
 \sqrt{\alpha \gamma \delta \over \beta} $$
and
$$e^{2 E } = {\alpha \beta \over \gamma \delta} \  \  ,$$
one sees that the system satisfies detailed balance for $\alpha',\gamma',\beta,\delta$.
Therefore the Gallavotti-Cohen symmetry implies  (\ref{Gallavotti-Cohen})  for the SSEP.
\newpage
\section{ The analogy with a multi-particle ruin  problem \label{ruin}}
In this appendix, we show  the similarity between  the  equations one has to solve
at each level of the hierarchy and   the equations which one can write  in a multi-particle ruin problem.

Consider first  a single particle which diffuses on  a chain of $N$ sites
with open boundary conditions. If the particle is at site $i$ at time
$t$, it jumps, during an  infinitesimal time interval $dt$, to site $i+1$
with probability $dt$ (for $ 1 \leq i \leq N-1$)  and   to site $i-1$
with probability $dt$  (for $2 \leq i \leq N$).
Moreover, a particle at  site  $1$  is absorbed at the left boundary with
probability $\alpha dt$
and  a particle at site $N$   is absorbed at the right boundary with  probability $\beta dt$.
In the usual ruin problem \cite{WF}, one asks the following question: what is the probability $T_i$  that a particle starting at site $i$ will escape at the left boundary.
Clearly $T_i$ satisfies for  $2 \leq i \leq N-1$
$$ T_{i+1} + T_{i-1} - 2 T_i =0$$
and at the boundaries
$$ \alpha + T_{2}  - (1 + \alpha)  T_1 =0$$
$$ T_{N-1}    - (1 + \beta)  T_N =0$$
These are precisely the  equations
(\ref{Ti1})-(\ref{TN1})
  we had to solve in section
4, if one takes $\gamma=\delta=0$ and $z=1$ (which implies that $\mu=0$
see (\ref{mu1})). 

The solution of this ruin problem is of course linear in $i$
\begin{equation}
 T_i = { N +{1 \over   \beta} - i  \over N + {1 \over \alpha} +{1 \over
\beta} -1}  \ .
\label{ruin1}
\end{equation}

Let us now generalize the ruin problem to two particles (the generalization to more particles is straightforward).
Consider two particles initially   at sites $i < j$ which  diffuse in the
same way as in the one-particle ruin problem, except that  the two
particles  are not allowed to occupy the same site. As time goes on, one
of the two particles will escape at one of the two boundaries, then the
other particle will diffuse until it also escapes.

Now we want to calculate the probability $U_{i,j}$ that both particles will escape through the left boundary.
One can write down the equations satisfied by $U_{i,j}$

\begin{equation}
  U_{i-1,j} + U_{i+1,j} + U_{i,j-1} +
U_{i,j+1} - 4 U_{i,j} =0 
\label{UU1}
\end{equation}
\begin{equation}
U_{i,i}+ U_{i+1,i+1} = 2 U_{i,i+1}
\label{UU2}
\end{equation}
\begin{equation}
 (3 + \alpha)  U_{1,i} = \alpha  T_i  + U_{2,i}+ U_{1,i+1} + U_{1,i-1}
\label{UU3}
\end{equation}
\begin{equation}
( 3+  \beta)  U_{i,N} = U_{i,N-1} + U_{i+1,N} + U_{i-1,N}
\label{UU4}
\end{equation}
 and they are identical to (\ref{Uij1})-(\ref{boundary4}) when
$\gamma=\delta=0$ and $z=1$ (implying that $\mu=0$).
It is not obvious a priori that the solution of these equations is simple.
However the  solution  turns out to be  linear in $i$ and $j$
\begin{equation}
U_{i,j}= {(N + {1 \over \beta} - 1 - i) (N  + {1 \over \beta}- j )
 \over (N  +{1 \over \alpha}  +{1 \over \beta} -1 ) (N  + {1 \over
\alpha} + {1 \over \beta} -2 )}  \ .
\label{ruin2}
\end{equation}
We also see in (\ref{ruin1}),(\ref{ruin2}) that the correlation between the two
particles is  weak
$$ u_{i,j} =  U_{i,j} - T_i T_j = O\left( {1 \over N} \right) $$
when $i$ and $j$ are of order $N$.
This weak correlation \cite{HS1}, which are similar to those seen in (\ref{scal2}), (\ref{scal3}), is however responsible for the non-Poissonian
character of the fluctuations of the integrated current.

Another quantity  which has a simple expression (i.e. for which the solution is linear in $i$ and $j$)
is the probability $U_{i,j}$ that
 one particle escapes at the right and the other particle escapes at the left, without specifying  on which side the first particle to escape leaves
 (starting with two particles at positions $i$ and $j$).
In this case the equations to solve are again  (\ref{UU1}),(\ref{UU2}) with boundary conditions (\ref{UU3}),(\ref{UU4})    replaced  by
\begin{equation}
 (3 + \alpha)  U_{1,i} =  \alpha (1 - T_i) + U_{2,i}+ U_{1,i+1} + U_{1,i-1}
\label{UU3ter}
\end{equation}
\begin{equation}
( 3+  \beta)  U_{i,N} =  \beta T_i + U_{i,N-1} + U_{i+1,N} + U_{i-1,N}
\label{UU4ter}
\end{equation}
and the solution  is
$$
U_{i,j}= { N(i+j) - 2 i j - 2 N +2 i + {1 \over \alpha} ( 2 N - i - j) + {1
\over \beta} ( i + j - 2) + {2 \over \alpha \beta} 
 \over (N - 1 +{1 \over \alpha}  +{1 \over \beta}) (N - 2 + {1 \over
\beta} + {1 \over \alpha})}  \ .
$$

If one asks however  a slightly  more precise question, namely what is the probability $U_{i,j}$   that (starting with a particle at $i$ and a particle at $j$), 
the first  particle  to escape leaves at the right boundary, and then the remaining particle escapes at the left boundary, 
the equations to solve are still  (\ref{UU1}),(\ref{UU2})  but with the boundary conditions  (\ref{UU3}),(\ref{UU4})  replaced by
\begin{equation}
 (3 + \alpha)  U_{1,i} =  U_{2,i}+ U_{1,i+1} + U_{1,i-1}
\label{UU3bis}
\end{equation}
\begin{equation}
( 3+  \beta)  U_{i,N} =  \beta T_i + U_{i,N-1} + U_{i+1,N} + U_{i-1,N} \ .
\label{UU4bis}
\end{equation}
These new boundary conditions make the problem much harder and one can check that the solution is no longer linear 
(or even polynomial) in $i$ and $j$.

So the same problem (\ref{UU1}),(\ref{UU2}) with the boundary
conditions (\ref{UU3}),(\ref{UU4}) or (\ref{UU3ter}),(\ref{UU4ter}) is
easy (the solution is linear in $i$ and $j$) whereas it is hard with boundary
conditions (\ref{UU3bis}),(\ref{UU4bis}). 
The main reason which made possible the calculation of the cumulants in the present paper is that
each time we had to solve equations of the type (\ref{UU1}),(\ref{UU2}),
the boundary conditions were such that the solution was polynomial in the
coordinates $i$ and $j$.
\newpage
\section{ The hierarchy \label{hierarchy}}
\subsection{The hierarchy for the correlation functions}
The hierarchy of section 4 can be summarized as follows:
\\ \\ \\ 
\underline{The equation for $\mu$}
\begin{equation}
\mu = \alpha ( z -1) -{(\alpha z + \gamma)(z-1)\over z} T_1
\label{mu2}  
\end{equation}
\underline{The  bulk equations}
\begin{equation}
\mu  T_i = \alpha ( z -1)T_i  -{(\alpha z + \gamma)(z-1)\over z}
 U_{1,i}
 + T_{i-1} + T_{i+1} - 2 T_i
\label{Ti2}  
\end{equation}
\begin{equation}
 \mu U_{i,j} = \alpha  (z-1) U_{i,j}  -{(\alpha z + \gamma)(z-1)\over z}
 V_{1,i,j}
 + U_{i-1,j} + U_{i+1,j} + U_{i,j-1} +
U_{i,j+1} - 4 U_{i,j}
\label{Uij2}  
\end{equation}
\begin{eqnarray}
 \mu V_{i,j,k} = \alpha  (z-1) V_{i,j,k}  -{(\alpha z + \gamma)(z-1)\over z}
 W_{1,i,j,k}
 + V_{i-1,j,k} + V_{i+1,j,k} 
\nonumber \\ + V_{i,j-1,k} +
V_{i,j+1,k } + V_{i,j,k-1} + V_{i,j,k+1}  - 6 V_{i,j,k}
\label{Vijik2}  
\end{eqnarray}
\underline{The left boundary equations}
\begin{equation}
 \alpha (z-1) T_1 -{(\alpha z + \gamma)(z-1)\over z}
 U_{1,1}  + T_0 - T_1  = \alpha
z - (\alpha z + \gamma) T_1
\label{T12}
\end{equation}
\begin{equation}
\alpha  (z-1) U_{1,i}  -{(\alpha z + \gamma)(z-1)\over z} V_{1,1,i}
 + U_{0,i} - U_{1,i} = \alpha z T_i -
 (\alpha z+ \gamma)U_{1,i}
\label{U1i2}
\end{equation}
\begin{equation}
\alpha  (z-1) V_{1,i,j}  -{(\alpha z + \gamma)(z-1)\over z} W_{1,1,i,j}
 + V_{0,i,j} - V_{1,i,j} = \alpha z U_{i,j} -
 (\alpha z+ \gamma)V_{1,i,j}
\label{V1ij2}
\end{equation}
\underline{The right boundary equations}
\begin{equation}
\delta - (\beta + \delta) T_N = T_{N+1} - T_N
\label{TN2}
\end{equation}
\begin{equation}
\delta T_i - (\beta + \delta) U_{i,N} = U_{i,N+1} - U_{i,N}
\label{UiN2}
\end{equation}
\begin{equation}
\delta U_{i,j} - (\beta + \delta) V_{i,j,N} = V_{i,j,N+1} - V_{i,j,N}
\label{VijN2}
\end{equation}
\underline{The equations for adjacent particles}
\begin{equation}
U_{i,i}+U_{i+1,i+1} = 2 U_{i,i+1}
\label{Uii2}
\end{equation}
\begin{equation}
V_{i,i,j}+V_{i+1,i+1,j} = 2 V_{i,i+1,j}
\label{Viij2}
\end{equation}
\begin{equation}
V_{i,j,j}+V_{i,j+1,j+1} = 2 V_{i,j,j+1}
\label{Vijj2}
\end{equation}
When one solves this hierarchy up to order $3$ in $\rho_a$ and $\rho_b$,
one obtains
{\small
\begin{eqnarray}
\mu= { (\rho_a z - \rho_b)(z-1) \over z N_1} 
+(z-1)^2 \left[
{(a - 3 a^2 + 2 a^3 + b - 3 b^2 + 2 b^3)   (\rho_a  z -
\rho_b)^2\over 6 z^2 N_1^3 (N_1-1)} \right.
\nonumber \\
\left.
-{ (\rho_a z - \rho_b)^2\over 6 z^2 N_1^2 (N_1-1)}
+{  \rho_a^2 z^2 +  \rho_b^2\over 2 z^2  N_1 (N_1-1)}
-  {  \rho_a^2 z^2 + \rho_a  \rho_b z  +  \rho_b^2\over 3 z^2  (N_1-1)}
\right]
\nonumber \\
+ (z-1)^3\left[
 {(a-3 a^2 + 2 a^3 + b - 3 b^2 + 2 b^3)^2  ( \rho_a  z - 
\rho_b)^3 \over 9 z^3 N_1^5 (N_1-1) (N_1 -2) } \right.
\nonumber \\
+{(-7 a + 30 a^2 - 50 a^3 + 45 a^4 - 18 a^5 - 7 b + 30 b^2 - 50 b^3 + 45
b^4 - 18 b^5 )(\rho_a  z - \rho_b)^3\over 45 z^3 N_1^4 (N_1-1) (N_1 -2) }
\nonumber \\
+{[(2 + 15 a - 45 a^2 + 30 a^3 + 15 b - 45 b^2 + 30 b^3) (\rho_a^2 z^2  +
\rho_b^2  ) - 4 \rho_a \rho_b z ](\rho_a z - \rho_b)\over 45 z^3 N_1^3
(N_1-1) (N_1 -2) }
\nonumber \\
 - {(a - 3 a^2 + 2 a^3 + b - 3 b^2 + 2 b^3)( \rho_a^3 z^3 - \rho_b^3) +3
(\rho_a^2 z^2 + \rho_b^2 )(\rho_a z - \rho_b)
\over 9 z^3 N_1^2 (N_1-1) (N_1 -2) } 
\nonumber \\
+ {7 (\rho_a^3 z^3 - \rho_b^3) \over 9 z^3 N_1 (N_1-1) (N_1 -2) }
- { (\rho_a z - \rho_b) ( 2 \rho_a^2 z^2 + 3 \rho_a  \rho_b z  + 2 \rho_b^2) \over 3 
  z^3  (N_1-1) (N_1 -2) }
\nonumber \\
\left. + {2 (\rho_a z - \rho_b) ( 4 \rho_a^2 z^2 + 7 \rho_a \rho_b z + 4 \rho_b^2)
N_1\over 45
z^3  (N_1-1) (N_1 -2)   } \right]
\nonumber \\
\label{mu3}
\end{eqnarray}
}
where $N_1= N+a+b-1$,  $\rho_a = \alpha /( \alpha+ \gamma)$,
 $\rho_b = \delta /( \beta+ \delta)$, $a= 1/ (\alpha+ \gamma)$ and
$b=1/( \beta+ \delta)$.

\subsection{The hierarchy for connected correlation   functions}
If one introduces the connected functions  $u_{i,j}$, $v_{i,j,k}$...
defined by
\begin{equation}
U_{i,j}= T_i T_j + u_{i,j} 
\label{uijdef}
\end{equation}
\begin{equation}
V_{i,j,k}= T_i T_j T_k + u_{i,j} T_k + u_{i,k} T_j + u_{j,k} T_i +
v_{i,j,k} 
\label{vijkdef}
\end{equation}
\begin{eqnarray}
W_{i,j,k,l}= T_i T_j T_k  T_l + u_{i,j} T_k T_l  + u_{i,k} T_j T_l  +
u_{j,k} T_i T_l  + u_{i,l} T_j T_k 
 \nonumber \\
 + u_{j,l} T_i T_k 
 + 
u_{k,l} T_i T_j + u_{i,j} u_{k,l} + u_{i,k} u_{j,l} + u_{i,l} u_{j,k}
+ v_{i,j,k} T_l 
 \nonumber \\
+ v_{i,j,l} T_k + v_{i,k,l} T_j 
+ v_{j,k,l} T_i +
w_{i,j,k,l} 
\label{conec3}
\end{eqnarray}
the hierarchy becomes
\\ \\ \\ 
\underline{The equation for $\mu$} (obtained by  combining (\ref{mu2}),
(\ref{Ti2}) and
(\ref{T12}))
\begin{equation}
\mu \left[ 1 - {z-1 \over z} T_1 \right]= {  z-1 \over z} (T_1- T_2)
\label{mup3}  
\end{equation}
\underline{The  bulk equations}
\begin{equation}
\epsilon  \  u_{1,i}
 = T_{i-1} + T_{i+1} - 2 T_i
\label{Ti3}  
\end{equation}
\begin{equation}
 \epsilon \ v_{1,i,j} = 
  u_{i-1,j} + u_{i+1,j} + u_{i,j-1} +
u_{i,j+1} - 4 u_{i,j}
\label{Uij3}  
\end{equation}
\begin{eqnarray}
 \epsilon \ w_{1,i,j,k}
 = v_{i-1,j,k} + v_{i+1,j,k} 
+ v_{i,j-1,k} +
v_{i,j+1,k } + v_{i,j,k-1} + v_{i,j,k+1}  - 6 v_{i,j,k}
\nonumber \\
\label{Vijk3}  
\end{eqnarray}
where $\epsilon $ is defined by
\begin{equation}
\epsilon ={(\alpha z + \gamma)( z-1) \over z}
\label{epsilondef}
\end{equation}
\underline{The left boundary equations}
\begin{equation}
( \alpha  z + \gamma)  T_1 -\alpha z 
  + T_0 - T_1  =  \epsilon \ u_{1,1} - \mu  \ T_1
\label{T13}
\end{equation}
\begin{equation}
(\alpha + \gamma) u_{1,i} + u_{0,i} - u_{1,i} = \epsilon  \  v_{1,1,i}  - 2 \mu \  u_{1,i}
\label{U1i3}
\end{equation}
\begin{equation}
(\alpha + \gamma) v_{1,i,j} + v_{0,i,j} - v_{1,i,j} = \epsilon  \  [w_{1,1,i,j} +
u_{1,i} u_{1,j}]   - 2 \mu \  v_{1,i,j}
\label{V1ij3}
\end{equation}
\underline{The right boundary equations}
\begin{equation}
\delta - (\beta + \delta) T_N = T_{N+1} - T_N
\label{TN3}
\end{equation}
\begin{equation}
 - (\beta + \delta) u_{i,N} = u_{i,N+1} - u_{i,N}
\label{UiN3}
\end{equation}
\begin{equation}
 - (\beta + \delta) v_{i,j,N} = v_{i,j,N+1} - v_{i,j,N}
\label{VijN3}
\end{equation}
\underline{The equations for adjacent particles}
\begin{equation}
u_{i,i}+u_{i+1,i+1} - 2 u_{i,i+1} = - (T_i - T_{i+1})^2
\label{Uii3}
\end{equation}
\begin{equation}
v_{i,i,j}+v_{i+1,i+1,j} - 2 v_{i,i+1,j} = -2 (T_i - T_{i+1}) ( u_{i,j} - u_{i+1,j})
\label{Viij3}
\end{equation}
\begin{equation}
v_{i,j,j}+v_{i,j+1,j+1} - 2 v_{i,j,j+1} = -2 (T_j - T_{j+1}) ( u_{i,j} -
u_{i,j+1})
\label{Vijj3}
\end{equation}
\begin{eqnarray}
w_{i,i,j,k}+w_{i+1,i+1,j,k} - 2 w_{i,i+1,j,k} =
 -2 (T_i - T_{i+1}) ( v_{i,j,k} - v_{i+1,j,k})
&& \nonumber \\
  -2 (u_{i,j} - u_{i+1,j}) ( u_{i,k} - u_{i+1,k}) &&
\label{Wiijk3}
\end{eqnarray}
etc... \\
(As already discussed right after (\ref{boundary4}), the (unphysical) values
$T_0, u_{i,i} ...$ are   obtained by
using the explicit (polynomial)  expressions of $T_{i}$,  $u_{i,j}$ 
for $i=0, j=i$).

\section*{Acknowledgments:}
B. Derrida thanks the hospitality of the Newton Institute,  Cambridge UK, in
the summer  2003,
where part of this work was done.


\newpage

\begin{thebibliography}{99}


     \bibitem{LLP} S. Lepri, R. Livi, A. Politi,
      Thermal conduction in classical low-dimensional lattices,
     {\it  Phys. Rep.} {\bf    377},  1-80 (2003)
     
\bibitem{BLR} F. Bonetto, J.L. Lebowitz, L. Rey-Bellet,
Fourier's law: a challenge to theorists, Mathematical Physics 2000, 128-150,
  2000, Imperial College Press,
      math-ph/0002052 
\bibitem{BeB}C. W. J. Beenakker M. B\"uttiker Suppression of shot noise in metallic diffusive conductors, {\it Phys. Rev. B}  {\bf 46}, 1889-1892 (1992) 

\bibitem{BlB} 
   Y.M. Blanter, M. B\"uttiker, 
     Shot noise in mesoscopic conductors, {\it Phys. Rep.} {\bf 336}, 1-166  (2000)

\bibitem{LLY}
     Hyunwoo Lee, L. S. Levitov, A. Yu. Yakovets, Universal statistics of transport in disordered conductors, {\it Phys. Rev. B} {\bf  51}, 4079-4083 (1995)

\bibitem{Naz} Yu. V. Nazarov,  Universality of weak localization
{\it  Ann. Phys.} (Leipzig) {\bf 8}, Spec. Issue, 193-197 (1999)
 (see also cond-mat/9908143)

\bibitem{GGM}  D.B. Gutman,Y.  Gefen and A. Mirlin, High cumulants of currents
 fluctuations out of equilibrium
 in {\it Proc. of Quantum Noise}, ed. Yu. V. Nazarov and Ya. M. Blanter
 (Kluwer, Dordreicht) under press
 (see also cond-mat/0210076 )

\bibitem{N0} K. E. Nagaev,
On the shot noise in dirty metal contacts, {\it Phys. Lett. A} {\bf 169}
103-107 (1992)
\bibitem{DB} 
M. J. M. de Jong, C. W. J. Beenakker, 
Semiclassical theory of shot-noise suppression,
{\it Phys. Rev. B} {\bf51}, 16867-16870 (1995)

\bibitem{N} K. E. Nagaev,
 Cascade Boltzmann-Langevin approach to higher-order current correlations in diffusive metal contacts,
{\it      Phys. Rev. B} {\bf 66}, 075334 (2002)


\bibitem{RD} P.-E. Roche, B. Dou\c{c}ot,
Shot-noise statistics in diffusive conductors,
{\it Eur. Phys. J. B} {\bf 27},  393-398 (2002)

\bibitem{GG} D. B. Gutman and Y. Gefen, 
Shot noise at high temperatures,
{\it Phys. Rev. B } {\bf 68}, 035302 (2003)


 \bibitem{Ligg} T. M. Liggett,   {\it Stochastic interacting systems:
contact, voter, and exclusion processes} (Springer-Verlag, New York, 1985)
 \bibitem{KOV} C.~Kipnis, S.~Olla, S.~R.~S.~Varadhan, Hydrodynamics and
large deviations for simple exclusion processes, {\it Commun. Pure
Appl. Math.}~{\bf 42}, 115--137 (1989)
 \bibitem{HS} H.~Spohn, {\it Large Scale Dynamics of Interacting Particles}
(Springer-Verlag, Berlin, 1991)
 \bibitem{ZS} B. Schmittman, R. K. P. Zia, {\it Statistical mechanics of
driven diffusive systems} (Academic Press, London, 1995)

 \bibitem{SS} J.E. Santos, G.M. Schutz,  
Exact time-dependent correlation functions for the symmetric exclusion
process with open boundary,
{\it Phys.  Rev. E} {\bf  64},  036107  (2001)


 \bibitem{DEHP}  B.~Derrida, M.~R.~Evans, V.~Hakim, V.~Pasquier,
Exact solution of a 1D asymmetric exclusion model using a matrix
formulation, {\it J.~Phys.~A}~{\bf 26},  1493--1517 (1993)


 \bibitem{DLS1} B.  Derrida, J.  L.  Lebowitz, E.  R.  Speer, Free energy
functional for nonequilibrium systems: an exactly solvable case,
{\it Phys. Rev. Lett.}~{\bf 87}, 150601(2001)

\bibitem{DLS2} B. Derrida, J.L. Lebowitz, E.R. Speer
{Large Deviation of the Density Profile in the Steady State of the Open
  Symmetric Simple  Exclusion Process}
{\it J. Stat. Phys.} {\bf 107}, 599 (2002)

 \bibitem{BDGJL} L.  Bertini, A.  De Sole, D.  Gabrielli, G.  Jona--Lasinio,
C.  Landim, Fluctuations in stationary non equilibrium states of
irreversible processes, {\it Phys.  Rev.  Lett.}~{\bf 87}, 040601 (2001) 

 \bibitem{BDGJL2} L.  Bertini, A.  De Sole, D.  Gabrielli, G.  Jona--Lasinio,
C.  Landim, Macroscopic fluctuation theory for stationary non equilibrium
states,  {\it J. Stat, Phys.} {\bf 107},  635-675  (2002)


\bibitem{R}P.M. Richards, Theory of one-dimensional hopping conductivity and diffusion, 
                                         { \it Phys. Rev. B}  {\bf 16}, 1393-1409 (1977)
\bibitem{BKT}H. van Beijeren, R. Kutner, H. Spohn, Excess Noise for Driven Diffusive Systems,  {\it Phys. Rev. Lett.} {\bf 54}, 2026-2029 (1985)
\bibitem{FF}  P.A. Ferrari, L. R. Fontes,
Current fluctuations for the asymmetric simple exclusion process, 
{\it Ann. Probab.} {\bf 22}, 2 820-832 (1994) 

\bibitem{DL} B. Derrida, J.L.  Lebowitz,
Exact large deviation function in the asymmetric
exclusion process,  {\it Phys. Rev. Lett.}  {\bf 80}, 209-213 (1998)
\bibitem{PS} M. Pr\"ahofer and H. Spohn, 
Current fluctuations for the totally asymmetric simple exclusion process, cond-mat/0101200 
in "In and out of equilibrium", ed. V. Sidoravicius, Progress in Probability, Birkh\a"user 2002 

\bibitem{LK} Deok-Sun Lee and D. Kim, Large deviation function of the partially asymmetric exclusion process
{\it Phys. Rev. E} { \bf 59}, 6476-6482 (1999) 

\bibitem{DA} B. Derrida, C. Appert,
 Universal large deviation function
of the Kardar-Parisi-Zhang equation in one dimension, {\it   J. Stat. Phys.}
{\bf 94}, 1-30  (1999)
\bibitem{PPH} A.M. Povolotsky, V.B. Priezzhev, C.K. Hu,
The asymmetric avalanche process,
{\it J. Stat. Phys.} {\bf 111}, 1149-1182 (2003) 

\bibitem{DEMal} B. Derrida, M.R. Evans, K. Mallick, Exact diffusion constant of a
one dimensional  asymmetric exclusion model with open boundaries, {\it J.
Stat. Phys. } {\bf 79}, 833-874 (1995)


 \bibitem{GC} 	 G. Gallavotti, E.D.G. Cohen, Dynamical ensembles in stationary states, {\it J. Stat. Phys.} {\bf 80}, 931-970 (1995)
 \bibitem{LS} 	J.L. Lebowitz, H.~Spohn, A Gallavotti-Cohen Type Symmetry in the Large Deviation Functional for Stochastic Dynamics
{\it J.   Stat. Phys.} {\bf 95}, 333-366 (1999)



 \bibitem{HS1} H.~Spohn, Long range correlations for stochastic lattice
gases in a non-equilibrium steady state, {\it J.~Phys A.} {\bf 16},
4275--4291 (1983)


 \bibitem{GLMS} P Garrido, J. L. Lebowitz. C, Maes, H. Spohn, Long-range
correlations for conservative dynamics, {\it Phys. Rev. A} {\bf 42},
1954--1968 (1990)


 \bibitem{Sas} T.~Sasamoto, One dimensional partially asymmetric simple
exclusion process with open boundaries: Orthogonal polynomials approach,
{\it J.~Phys.~A}~{\bf 32}, 7109--7131 (1999)

 \bibitem{BECE} R.~A.~Blythe, M.~R.~Evans, F.~Colaiori, F.~H.~L.~Essler,
Exact solution of a partially asymmetric exclusion model using a deformed
oscillator algebra, {\it J.~Phys.~A}~{\bf 33}, 2313--2332 (2000)

\bibitem{ED}
 C. Enaud, B. Derrida,  Large deviation functional of the weakly
asymmetric exclusion process, submitted to  {\it J. Stat. Phys.}, cond-mat/0307023 

 
\bibitem{WF}
W. Feller,
{\it  An introduction to probability theory and its applications},
 (Wiley, New York 1968) 
\end{thebibliography}
\end{document}